\documentclass[amsmat,amssymb,amsfonts,aps,prl,twocolumn,showpacs]{revtex4}

\usepackage{graphicx}
\usepackage{dcolumn}
\usepackage{bm}
\begin{document}

\title{Kondo effect and conductance of nanocontacts with magnetic impurities}

\author{ D. Jacob}
\email{djacob@physics.rutgers.edu}
\affiliation{Dept. of Physics \& Astronomy, Rutgers University, 136 Frelinghuysen Rd., Piscataway, NJ-08854}
\author{ K. Haule and G. Kotliar}
\affiliation{Dept. of Physics \& Astronomy, Rutgers University, 136 Frelinghuysen Rd., Piscataway, NJ-08854}

\date{\today}


\begin{abstract}
  We study the impact of dynamical correlations on the electronic structure and coherent transport 
  properties of Cu nanocontacts hosting a single magnetic impurity (Ni,Co,Fe) in the contact region.
  The strong dynamical correlations of the impurity $3d$-electrons are fully taken into account by 
  combining density functional calculations with a dynamical treatment of the impurity $3d$-shell 
  in the one-crossing approximation. We find that dynamical correlations give rise to the Kondo effect 
  and lead to Fano features in the coherent transport characteristics similar to those observed in 
  related experiments. 
\end{abstract}

\maketitle


The development of nanoscale spintronics devices based on single molecules
and atomic size junctions containing magnetic atoms is a fascinating and challenging field of 
research at the moment 
\cite{Tsukagoshi-Zhao-Hueso-Bogani}.
An important contribution to the electronic structure and transport properties of these devices
comes from the strongly interacting $d$- or $f$-electrons of the magnetic atoms. The strong interactions 
result in dynamical correlations that give rise to interesting effects like e.g. the Kondo effect.
For example, Fano lineshapes \cite{Fano:pr:61} observed in the conductance characteristics of scanning 
tunneling microscope (STM) experiments with magnetic adatoms and molecular complexes on metal surfaces 
\cite{Madhavan-Vitali-Neel,Neel:prl:07} are the result of Kondo resonances at the Fermi level 
\cite{Kondo:ptp:64,Schiller:prb:00}. 
Recently, Fano lineshapes have also been observed in the non-linear conductance characteristics 
of chemically homogeneous nanocontacts \cite{Agrait:pr:03} made from ferromagnetic transition 
metals (Ni, Co, Fe) \cite{Calvo:nature:09}. 

State of the art calculations of the conductance and current through atomic- and molecular-size
conductors consist in combining {\it ab initio} electronic structure calculations on the level of 
density-functional theory (DFT) with the non-equilibrium Green's function (NEGF) technique 
\cite{abinitnegf}. This methodology works quite well for metallic nanocontacts \cite{Agrait:pr:03} 
predicting zero-bias conductances that are in general in good agreement with experiments 
\cite{nanocontacts}.
However, static mean-field methods like DFT cannot describe dynamical electron correlations. 
Thus the DFT based {\it ab initio} transport methodology is not capable of describing the 
Fano-Kondo lineshapes \cite{Fano:pr:61} observed in STM studies of magnetic adatoms 
on surfaces \cite{Madhavan-Vitali-Neel,Neel:prl:07}.

In order to explore the impact of strong dynamic correlations on the transport properties
of atomic- and molecular-size conductors, we study Cu nanocontacts hosting magnetic impurities 
in the contact region. Such a system could also be realized experimentally with e.g. the break 
junction technique \cite{Agrait:pr:03} using alloys containing magnetic atoms like e.g. Cupronickel. 
To this end we extend the established DFT based {\it ab initio} quantum transport methodology 
to incorporate dynamic electron correlations by adapting the LDA+DMFT method \cite{Kotliar:rmp:06} to 
the case of a single magnetic impurity in a nanocontact. While the strong dynamic correlations of the 
impurity $d$-electrons are fully taken into account, the rest of the system is described on a static 
mean-field level in the local density approximation (LDA) to DFT. 
Other recent approaches to include dynamic electron correlations in the {\it ab initio} description 
of quantum transport are based on the GW approximation (GWA) \cite{Thygesen:jcp:07} or the three-body
scattering formalism (3BS) \cite{Ferretti:prl:05}. 
While the GWA is only suitable for weakly correlated systems due to the perturbative treatment of the 
electron-electron interactions, the 3BS is in principle capable of describing more strongly correlated 
systems as it goes beyond perturbation theory.
However, the 3BS does not provide a satisfactory solution of the Anderson impurity problem since the local 
correlations are not taken into account properly. In contrast, in our method both the strong Coulomb interactions 
between the impurity $d$-electrons and the resulting local correlations are taken into account properly. 

\begin{figure}
  \begin{tabular}{cc}
    \includegraphics[width=0.45\linewidth]{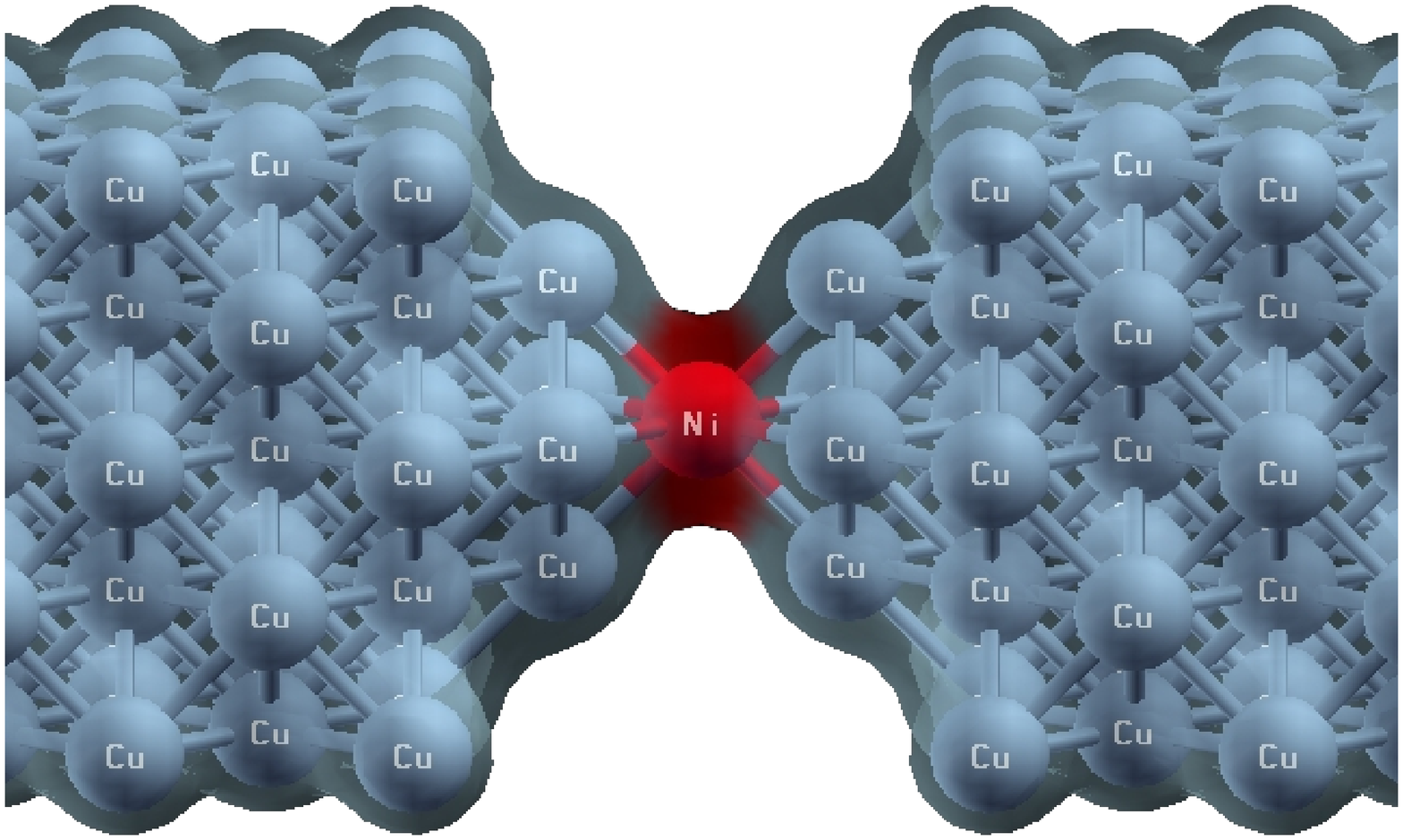}&
    \includegraphics[width=0.45\linewidth]{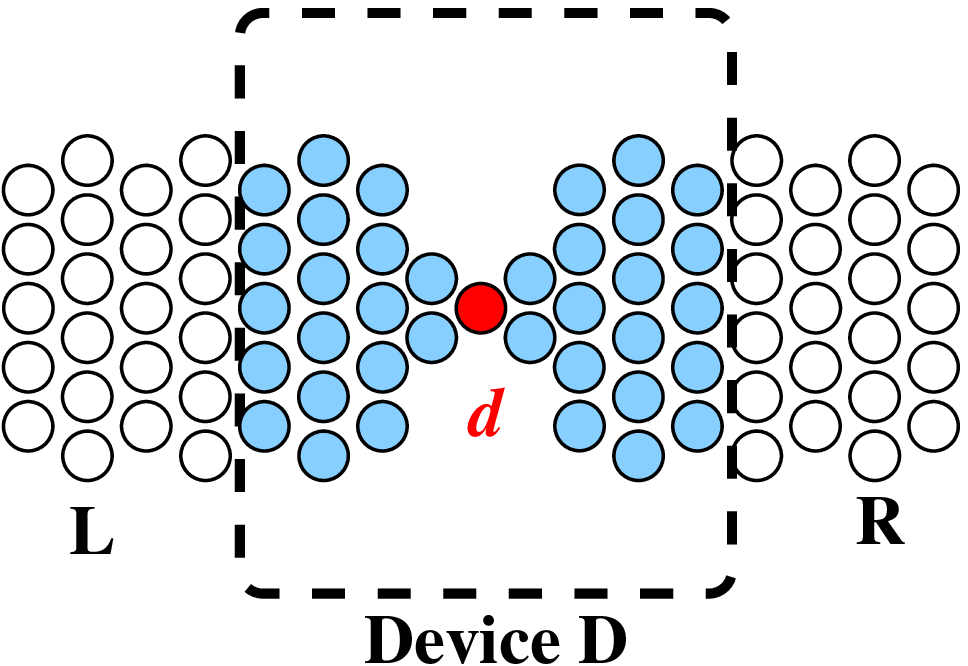} 
  \end{tabular}
  \caption{(Color online)
    Left: Atomic model of a Cu nanocontact with a magnetic 
    impurity (red) in the contact region. Right: Division of system 
    into left (L) and right (R) electrode, and  central device region (D)
    containing the magnetic impurity (red) hosting the strongly correlated 
    $d$-orbitals.
  }
  \label{fig:model}
\end{figure}

We consider a single magnetic impurity bridging the tips of two semi-infinite 
Cu nanowires of finite width grown in the (001) direction as shown in Fig. \ref{fig:model}. 
We divide the system into three parts as shown in the right panel of Fig. \ref{fig:model}: 
Two semi-infnite leads L and R, and the central device region (D) which contains the central
magnetic impurity with the strongly interacting $3d$-shell ($d$), and the tips of the two 
electrodes. The device also contains a sufficient part of the semi-infinite
leads so that the two leads L and R are sufficiently far away from the scattering region and
the electronic structure of the leads has relaxed to that of bulk (i.e. infinite) nanowires.
The effective one-body Hamiltonians of the device region and leads are obtained from DFT 
calculations on the level of LDA. Here we use the supercell approach \cite{epaps} to obtain the 
effective Kohn-Sham (KS) Hamiltonians of each part of the system prior to the dynamical treatment 
of the impurity $d$-shell and the transport calculations. The electronic structure of the device 
region is calculated with the CRYSTAL06 {\it ab initio} electronic structure program for periodic 
systems \cite{Crystal:06} by definining a one-dimensional periodic system consisting of the device 
region as the unit cell. 
The device Hamiltonian $\mathbf{H}_D$ 
is then obtained from the converged KS Hamiltonian of the unit cell of the periodic system. In the same 
way, the unit cell Hamiltonians $\mathbf{H}^0_{\rm L/R}$ and hoppings $\mathbf{V}_{\rm L/R}$ between unit 
cells of the left and right leads can be extracted from calulcations of infinite nanowires with finite
width since the electronic structure in the semi-infinite leads has relaxed to that of an infinite 
nanowire. In the LDA calculations we employ a minimal basis set plus effective core pseudo-potential 
that takes into account only the $4s$, $4p$ and $3d$ valence shells of the Cu atoms and the magnetic 
impurity \cite{Hurley:jcp:86}.

The strong electron correlations in the $3d$-shell of the magnetic impurity are captured 
by adding a Hubbard-like interaction term to the one-body Hamiltonian within the correlated 
subspace $d$:
$\hat{\mathcal H}_U = \frac{1}{2}\sum U_{ijkl} \,
\hat{c}_{i\sigma_1}^\dagger\hat{c}_{j\sigma_2}^\dagger \hat{c}_{l\sigma_2}\hat{c}_{k\sigma_1}$
(Einstein sum convention).
$U_{ijkl}$ are the matrix elements of the effective Coulomb interaction of 
the $3d$-electrons which is smaller than the bare Coulomb interaction due 
to the screening by the conduction electrons. In the spherical approximation 
all matrix elements $U_{ijkl}$ can be calculated from the Slater integrals
$F^0$, $F^2$, and $F^4$ which are related to the average Coulomb repulsion
$U$ between electrons and to the Hund's rule coupling $J$ by $F^0=U$, $F^2=(14/1.625)J$, 
and $F^4=0.625F^4$ \cite{Kotliar:rmp:06}. For $3d$ transition metal elements in bulk 
materials the repulsion $U$ is around 2-3~eV and $J$ is around 1~eV \cite{Grechnev:prb:07}. 
Due to the lower coordination of the contact atoms the screening of the direct interaction 
is reduced compared to its bulk value. Here we take $U=5$~eV and $J=1$~eV, but we have 
checked that the results do not change much when $U$ is varied between 4 and 6~eV.

The Coulomb interaction within the correlated $3d$ subspace has already been taken 
into account on a static mean-field level in the effective KS Hamiltonian of the 
device. Therefore the KS Hamiltonian within the correlated subspace $\mathbf{H}_d$ 
has to be corrected by a double-counting correction term, i.e. 
$\mathbf{H}_{d}\equiv\mathbf{H}_{d}^{\rm KS}-\mathbf{H}_{\rm dc}$.
Here we use the standard expression, 
$\mathbf{H}_{\rm dc} = [U(N_d-\frac{1}{2})-\frac{1}{2}J(N_d-1)]\times\mathbf{I}_d$
where $\mathbf{I}_d$ is the identity matrix in the $d$ subspace, and $N_d$ is the 
occupation of the impurity $3d$-shell \cite{Kotliar:rmp:06}.

The central quantity is the Green's function (GF) of the device region:
\begin{equation}
  \label{eq:GD}
  \mathbf{G}_{\rm D} = \left(\omega+\mu-\mathbf{H}_{\rm D}
    +\mathbf{H}_{\rm dc}
    -\mathbf\Sigma_d-\mathbf\Sigma_{\rm L}
    -\mathbf\Sigma_{\rm R}\right)^{-1}
\end{equation}
where $\mu$ is the chemical potential.
$\mathbf\Sigma_{\rm L}$ and $\mathbf\Sigma_{\rm R}$ 
are self-energies that describe the coupling of the device to the semi-infinite 
leads L and R, respectively. 
These can be calculated from the effective one-body Hamiltonians of the leads by 
iteratively solving the Dyson equation
$\mathbf\Sigma_{\rm L/R}=\mathbf{V}_{\rm L/R}(\omega+\mu-\mathbf{H}^0_{\rm L/R}-
\mathbf\Sigma_{\rm L/R})^{-1}\mathbf{V}_{\rm L/R}^\dagger$. 
$\mathbf\Sigma_d$ is the local electronic self-energy that describes 
the dynamic electron correlations of the impurity $3d$-electrons.
In order to calculate $\mathbf\Sigma_d$, the generalized Anderson impurity 
problem given by the impurity $3d$-shell has to be solved. 
The impurity problem is described by the projection $\mathbf{P}_d$ of the GF (\ref{eq:GD}) 
onto the correlated subspace $d$: $\mathbf{G}_d\equiv\mathbf{P}_d \mathbf{G}_{\rm D} \mathbf{P}_d$
which can be written as 
\begin{equation}
  \label{eq:Gd}
  \mathbf{G}_d(\omega) = \left(\omega+\mu-\mathbf{H}_d-\mathbf\Sigma_d(\omega)-\mathbf\Delta_d(\omega)\right)^{-1}
\end{equation}
where we have introduced the so-called hybridization function $\mathbf\Delta_d$ 
which describes the hybridization of the impurity electrons with the 
conduction electrons. The hybridization function can be calculated 
from the projection $\mathbf{P}_d$ of the \emph{uncorrelated} device GF 
$\mathbf{G}_D^0=\left(\omega+\mu-\mathbf{H}_D+\mathbf{H}_{dc}-\mathbf\Sigma_L-\mathbf\Sigma_R\right)^{-1}$
onto the correlated subspace $d$ \cite{epaps}, i.e. from
$\mathbf{G}_d^0\equiv \mathbf{P}_d \mathbf{G}_D^0 \mathbf{P}_d $.
Solving eq. (\ref{eq:Gd}) for $\mathbf\Delta_d$ and using
that $[\mathbf{G}_d]^{-1}=[\mathbf{G}_d^0]^{-1}-\mathbf\Sigma_d$
we obtain:
\begin{equation}
  \label{eq:Delta}
  \mathbf\Delta_d(\omega) = \omega+\mu-\mathbf{H}_d-[\mathbf{G}_d^0(\omega)]^{-1}
\end{equation}

The hybridization function $\mathbf\Delta_d$ \cite{epaps}, the Coulomb repulsion
$U$, the Hund's rule coupling $J$, and the impurity levels $\epsilon_{d,i}=(\mathbf{H}_d)_{ii}$ 
are the relevant parameters for solving the impurity problem. 
Here we employ the so-called One-Crossing-Approximation (OCA) to solve the impurity 
problem \cite{Haule:prb:01,Kotliar:rmp:06} which is particularly well 
suited for the Kondo regime.

The current through a strongly interacting impurity can be calculated 
exactly by the Meir-Wingreen formula \cite{Meir:prl:92}. However, for 
low temperatures and small bias voltages this expression is well 
approximated by the much simpler Landauer formula \cite{Landauer:philmag:70}: 
$ I(V) = \frac{2e}{h}\times\int_{0}^{eV}d\omega \, T(\omega) $
where $T(\omega)$ is the Landauer transmission function and where we have 
assumed an asymmetric voltage drop $V$ about the device region \cite{foot1}.
Thus the conductance is simply given by the Landauer transmission function:
$\mathcal{G}(V)=\frac{\partial I}{\partial V}(V) = \frac{2e^2}{h}\times T(eV)$.
The latter can be calculated from the device Green's function:
$T(\omega) = {\rm Tr}[ \mathbf\Gamma_L(\omega) \mathbf{G}_D^\dagger(\omega) 
  \mathbf\Gamma_R(\omega) \mathbf{G}_D(\omega) ]$
where $\mathbf\Gamma_{L/R}$ are the so-called coupling matrices which describe
the coupling to the leads, and can be calculated from the lead self-energies by
$\mathbf\Gamma_{L/R} = i ( \mathbf\Sigma_{L/R} - \mathbf\Sigma^\dagger_{L/R} )$.


\begin{figure}
  \includegraphics[width=\linewidth]{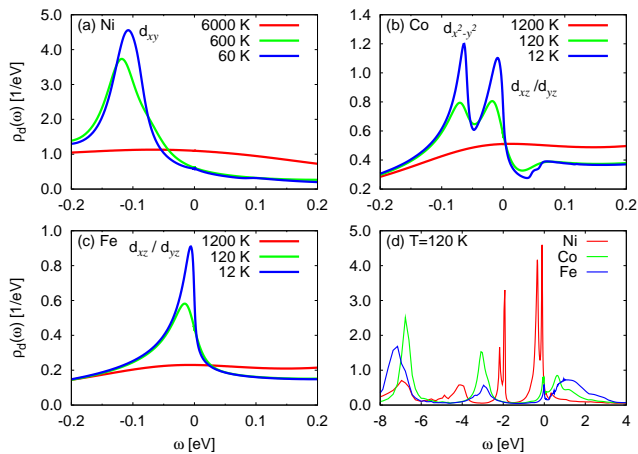}
  \caption{(Color online)
    PDOS of $d$-orbitals calculated with the LDA+OCA method 
    for different magnetic impurities in Cu nanocontact for 
    the geometry shown in Fig. \ref{fig:model}.
    (a)-(c) PDOS near Fermi level for a Ni (a), Co (b) and Fe 
    (c) impurity at different temperatures. 
    (d) Comparison of the PDOS of the three impurities on a 
    larger energy scale at $T=120$~K. 
    $U=5$~eV and $J=1$~eV in all cases.
  }
  \label{fig:pdos}
\end{figure}

\begin{figure}
  \includegraphics[width=\linewidth]{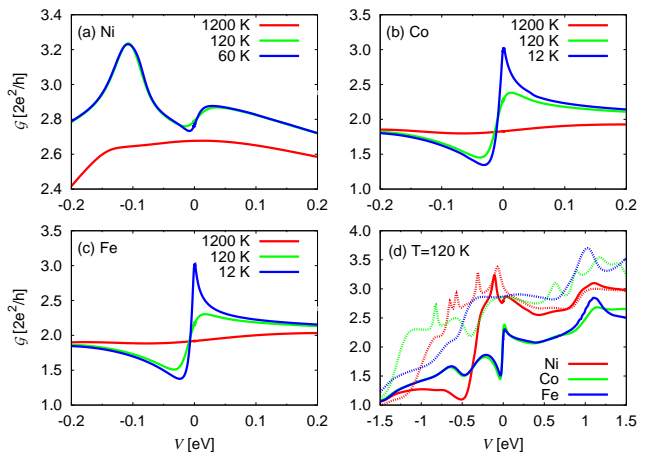}
  \caption{(Color online)
    Conductance calculated with LDA+OCA method 
    for different magnetic impurities in Cu nanocontact for
    the geometry shown in Fig. \ref{fig:model}:
    (a)-(c) Conductance $\mathcal{G}$ vs. bias voltage $V$
    for different temperatures and for small bias. 
    (d) Comparison of LDA+OCA (solid lines) and LSDA (dashed lines) 
    conductances for different impurities at $T=120$~K.
    $U=5$~eV and $J=1$~eV in all cases.
  }
  \label{fig:conductance}
\end{figure}

Fig. \ref{fig:pdos} shows the result of our LDA+OCA calculations for three 
different impurities: Ni, Co, and Fe, for different temperatures.
In all three cases the partial density of states (PDOS) for the impurity $3d$-electrons
shows resonances near the Fermi level which are temperature-dependent. More precisely,
the resonances vanish with increasing temperature. This is characteristic for the Kondo
effect which is usually observed only at low temperature. The resonances originate 
from different $d$-orbitals in each case, as indicated by the labels in the figures. 
In the case of  Ni, the resonance originates from the $d_{xy}$-orbital.
In the case of Co, there are two distinct peaks corresponding to 
two different sets of orbitals: The peak that is farther from (closer to) the 
Fermi level originates from the $d_{x^2-y^2}$ ($d_{xz},d_{yz}$) orbital(s).
The doubly-degenerate $d_{xz},d_{yz}$-orbitals are also responsible for the 
resonance in the case of Fe. 
As can be seen from Figs. \ref{fig:conductance}(a)-(c) the 
corresponding conductances all show Fano-like features. Interestingly,
in the case of Co, the $d_{x^2-y^2}$-resonance does not lead to a corresponding feature
in the conductance characteristics. This can be understood by the so-called 
{\it orbital blocking}: The electron transport through certain orbitals can be inhibited 
by the geometry or symmetry of atomic-size conductors in spite of the orbital having
spectral weight near the Fermi energy \cite{Jacob:prb:05}.

Tab. \ref{tab:occ} shows the orbital occupations and effective energy 
levels $\tilde\epsilon_d\equiv\epsilon_d+{\rm Re}\;\Delta_d(0)$ of the 
impurity $3d$-shell for different impurity atoms. One can see, that the 
effective energy levels $\tilde\epsilon_d$ roughly correlate with the orbital 
occupations except in the case of the $d_{x^2-y^2}$-orbital in Ni and Fe. 
To fully understand the orbital occupations, the imaginary part of the 
hybridization function $\Delta_d$ has to be taken into account. The 
imaginary part of $\Delta_d$ describes the broadening of the $d$-orbitals 
due to the coupling to the conduction electrons. It turns out that the 
$d_{x^2-y^2}$-orbital is by far the most localized orbital of all the 
$d$-orbitals i.e. the broadening is very small compared to the other 
$d$-levels \cite{epaps}. This explains why the $d_{x^2-y^2}$-orbital 
in Ni and Fe is only half-filled despite its low effective energy 
$\tilde\epsilon_d$.

Furthermore, we see that for Ni the $d_{xy}$-orbital which gives rise to the 
resonance near the Fermi energy is almost completely filled. Hence it does 
not carry a spin-1/2 and therefore the system is not in the Kondo regime, but 
is in the so-called {\it empty orbital} regime \cite{Hewson:book}. A broad 
quasiparticle (QP) peak quite close to the Fermi level appears which 
---as in the Kondo regime--- arises 
from the Fermi liquid behaviour at low temperatures. The temperature dependence 
of the QP peak is qualitatively similar to the Kondo regime, i.e. the peak broadens 
with increasing temperature and disappears above a critical temperature, called 
the {\it Kondo temperature} $T_{\rm K}$. 
For the Co and the Fe impurity, the doubly-degenerate orbitals $d_{xz}$ and $d_{yz}$ 
that give rise to the QP resonance at the Fermi level are occupied by three electrons 
(filling 3/4) and thus carry a spin-1/2. Therefore in the case of Co and Fe, the
system really is in the Kondo regime. 

We have estimated the Kondo temperatures $T_{\rm K}$ displayed in Tab. \ref{tab:occ} 
from the width of the QP peaks at low temperature. The Kondo temperatures 
follow the same trend, namely $T_{\rm K}({\rm Ni})>T_{\rm K}({\rm Co})>T_{\rm K}({\rm Fe})$,
which is also observed in STM experiments with adatoms on metal surfaces 
\cite{Jamneala:prb:00} and in pure transition metal nanocontacts \cite{Calvo:nature:09}.
Moreover, the Kondo temperature for Co agrees quite well with the $T_K$ estimated from 
recent STM experiments with Co adatoms on Cu surfaces in the {\it contact regime} 
\cite{Neel:prl:07}. 
The high Kondo temperatures of about 200~K for Co and Fe imply a strong antiferromagnetic coupling 
$J_{sd}$ between the conduction electrons and the impurity $d$-electrons giving rise to the Kondo 
effect since $T_{\rm K}\propto\exp(-1/J_{sd}\rho_0)$ where $\rho_0$ is the conduction electron 
DOS at the Fermi level \cite{Hewson:book}. 
This strong antiferromagnetic coupling might explain why the Kondo effect is observed
in ferromagnetic nanocontacts despite the ferromagnetic coupling to the bulk electrodes 
\cite{Calvo:nature:09}.

\begin{table}
  \setlength{\tabcolsep}{1mm}
  \begin{tabular}{|c|c|c|c|c|c|c|}
    \hline
    Imp.:          
    & \multicolumn{2}{c|}{Ni}
    & \multicolumn{2}{c|}{Co}
    & \multicolumn{2}{c|}{Fe} \\
    \hline
    & $n_d$ & $\tilde\epsilon_d$ [eV] 
    & $n_d$ & $\tilde\epsilon_d$ [eV] 
    & $n_d$ & $\tilde\epsilon_d$ [eV] \\
    \hline 
    $d_{3z^2-r^2}$  & 1.92 &  0               & 1.01 &            0    & 1.00 &            0    \\
    $d_{xz},d_{yz}$ & 3.60 & -0.22            & 3.05 &           -0.35 & 3.02 &           -0.71 \\
    $d_{x^2-y^2}$   & 1.00 & \phantom{-}0.05  & 1.98 &           -0.35 & 1.00 &           -0.43 \\
    $d_{xy }  $    &  1.90 & \phantom{-}0.26  & 1.00 & \phantom{-}0.03 & 1.00 & \phantom{-}0.24 \\ 
    \hline
    $N_d$
    & \multicolumn{2}{c|}{8.42}          
    & \multicolumn{2}{c|}{7.04}       
    & \multicolumn{2}{c|}{6.02} \\
    \hline
    $T_{\rm K}$ [K]       
    & \multicolumn{2}{c|}{$\sim$~500}      
    & \multicolumn{2}{c|}{$\sim$~225}  
    & \multicolumn{2}{c|}{$\sim$~160} \\
    \hline
  \end{tabular}
  \caption{
    Orbital occupations $n_d$, effective energy levels $\tilde\epsilon_d=\epsilon_d+{\rm Re}\;\Delta_d(0)$ 
    of impurity $d$-levels relative to $d_{3z^2-r^2}$-level, total occupation $N_d$ of the impurity $3d$-shell 
    and Kondo temperature $T_{\rm K}$ estimated from the width of the resonance closest to the Fermi level for 
    each of the three impurities. $U=5$~eV, $J=1$~eV.
  }
  \label{tab:occ}
\end{table}

Finally, in Fig. \ref{fig:conductance}(d) we compare the results obtained with the LDA+OCA 
method at low temperature with results obtained from DFT calculations on the level of the 
local spin density approximation (LSDA). For Ni (red lines) the effect of including 
dynamic correlations is only moderate. 
Thus for Ni the static mean-field description given by LSDA is a reasonable approximation 
to the fully correlated description by the LDA+OCA method, but at the cost of breaking the 
spin symmetry. This can be understood by recognizing that LSDA usually gives reasonable 
spectra for the empty orbital and mixed valence regimes.
In contrast for Co (green lines) and Fe (blue lines) taking into account dynamical correlations 
changes the conductance substantially. 
The dynamic correlations open a gap in the $3d$-shell thereby taking away spectral
weight from the Fermi level, leaving only the Kondo resonance (with small spectral weight)
at low temperatures. Consequently, the conductances predicted by LDA+OCA are considerably lower 
than the conductances predicted by LDA and LSDA.


In conclusion, we have extended the established DFT based {\it ab initio} 
transport methodology for nanoscopic conductors to include dynamic electron 
correlations. We find that nanocontacts hosting a magnetic impurity
show strong dynamical correlations which give rise to quasiparticle resonances 
at the Fermi level and corresponding Fano features in the conductance-voltage 
characteristics. 
Our findings agree well with experiments measuring the conductance 
through Co adatoms on metal surfaces in the contact regime.
Moreover, our results shed some light on the 
recent observation of the Kondo effect in ferromagnetic nanocontacts.

DJ acknowledges funding by the German academic exchange service (DAAD)
and fruitful discussions with J.~J. Palacios, J. Fern\'andez-Rossier, 
C. Untiedt and R. Calvo. KH was supported by the NSF under grant No. 
DMR 0746395 and by a Sloan fellowship. GK acknowledges funding by NSF 
under grant No. DMR 0528969.


\bibliographystyle{apsrev}
\bibliography{matcon,nano,theory,footnotes}

\begin{thebibliography}{32}
\expandafter\ifx\csname natexlab\endcsname\relax\def\natexlab#1{#1}\fi
\expandafter\ifx\csname bibnamefont\endcsname\relax
  \def\bibnamefont#1{#1}\fi
\expandafter\ifx\csname bibfnamefont\endcsname\relax
  \def\bibfnamefont#1{#1}\fi
\expandafter\ifx\csname citenamefont\endcsname\relax
  \def\citenamefont#1{#1}\fi
\expandafter\ifx\csname url\endcsname\relax
  \def\url#1{\texttt{#1}}\fi
\expandafter\ifx\csname urlprefix\endcsname\relax\def\urlprefix{URL }\fi
\providecommand{\bibinfo}[2]{#2}
\providecommand{\eprint}[2][]{\url{#2}}

\bibitem{Tsukagoshi-Zhao-Hueso-Bogani}
\bibinfo{author}{\bibfnamefont{K.}~\bibnamefont{Tsukagoshi}},
  \bibinfo{author}{\bibfnamefont{B.~W.} \bibnamefont{Alpenhaar}},
  \bibnamefont{and} \bibinfo{author}{\bibfnamefont{H.}~\bibnamefont{Ago}},
  \bibinfo{journal}{Nature} \textbf{\bibinfo{volume}{401}},
  \bibinfo{pages}{572} (\bibinfo{year}{1999});
\bibinfo{author}{\bibfnamefont{A.}~\bibnamefont{Zhao}} {\it et~al.},
  \bibinfo{journal}{Science}
  \textbf{\bibinfo{volume}{309}}, \bibinfo{pages}{1542} (\bibinfo{year}{2005});
\bibinfo{author}{\bibfnamefont{L.~E.} \bibnamefont{Hueso}} {\it et~al.},
  \bibinfo{journal}{Nature} \textbf{\bibinfo{volume}{445}},
  \bibinfo{pages}{410} (\bibinfo{year}{2007});
\bibinfo{author}{\bibfnamefont{L.}~\bibnamefont{Bogani}} \bibnamefont{and}
  \bibinfo{author}{\bibfnamefont{W.}~\bibnamefont{Wernsdorfer}},
  \bibinfo{journal}{Nature Materials} \textbf{\bibinfo{volume}{7}},
  \bibinfo{pages}{179} (\bibinfo{year}{2008}).

\bibitem[{\citenamefont{Fano}(1961)}]{Fano:pr:61}
\bibinfo{author}{\bibfnamefont{U.}~\bibnamefont{Fano}}, \bibinfo{journal}{Phys.
  Rev.} \textbf{\bibinfo{volume}{124}}, \bibinfo{pages}{1866}
  (\bibinfo{year}{1961}).

\bibitem{Madhavan-Vitali-Neel}
\bibinfo{author}{\bibfnamefont{V.}~\bibnamefont{Madhavan}} {\it et~al.},
\bibinfo{journal}{Science}
  \textbf{\bibinfo{volume}{280}}, \bibinfo{pages}{567} (\bibinfo{year}{1998});
  \bibinfo{author}{\bibfnamefont{L.}~\bibnamefont{Vitali}} {\it et~al.},
  \bibinfo{journal}{Phys. Rev. Lett.} \textbf{\bibinfo{volume}{101}},
  \bibinfo{pages}{216802} (\bibinfo{year}{2008});
  \bibinfo{author}{\bibfnamefont{N.}~\bibnamefont{Ne\'el}} {\it et~al.},
  \bibinfo{journal}{Phys. Rev. Lett.}
  \textbf{\bibinfo{volume}{101}}, \bibinfo{pages}{266803}
  (\bibinfo{year}{2008}).

\bibitem[{\citenamefont{N\'eel et~al.}(2007)\citenamefont{N\'eel, Kr\"oger,
  Limot, Palotas, Hofer, and Berndt}}]{Neel:prl:07}
\bibinfo{author}{\bibfnamefont{N.}~\bibnamefont{N\'eel}} {\it et~al.},
  \bibinfo{journal}{Phys. Rev. Lett.} \textbf{\bibinfo{volume}{98}},
  \bibinfo{pages}{016801} (\bibinfo{year}{2007}).

\bibitem[{\citenamefont{Kondo}(1964)}]{Kondo:ptp:64}
\bibinfo{author}{\bibfnamefont{J.}~\bibnamefont{Kondo}},
  \bibinfo{journal}{Prog. Theor. Phys.} \textbf{\bibinfo{volume}{32}},
  \bibinfo{pages}{37} (\bibinfo{year}{1964}).

\bibitem[{\citenamefont{Schiller and Hershfield}(2000)}]{Schiller:prb:00}
\bibinfo{author}{\bibfnamefont{A.}~\bibnamefont{Schiller}} \bibnamefont{and}
  \bibinfo{author}{\bibfnamefont{S.}~\bibnamefont{Hershfield}},
  \bibinfo{journal}{Phys. Rev. B} \textbf{\bibinfo{volume}{61}},
  \bibinfo{pages}{9036} (\bibinfo{year}{2000}).

\bibitem[{\citenamefont{Agra\"{\i}t et~al.}(2003)\citenamefont{Agra\"{\i}t,
  Yeyati, and van Ruitenbeek}}]{Agrait:pr:03}
\bibinfo{author}{\bibfnamefont{N.}~\bibnamefont{Agra\"{\i}t}},
  \bibinfo{author}{\bibfnamefont{A.~L.} \bibnamefont{Yeyati}},
  \bibnamefont{and} \bibinfo{author}{\bibfnamefont{J.~M.} \bibnamefont{van
  Ruitenbeek}}, \bibinfo{journal}{Physics Reports}
  \textbf{\bibinfo{volume}{377}}, \bibinfo{pages}{81} (\bibinfo{year}{2003}),
  \bibinfo{note}{and references therein}.

\bibitem[{\citenamefont{Calvo et~al.}(2009)\citenamefont{Calvo,
  Fern\'andez-Rossier, Palacios, Jacob, Natelson, and Untiedt}}]{Calvo:nature:09}
\bibinfo{author}{\bibfnamefont{M.~R.} \bibnamefont{Calvo}} {\it et~al.},
\bibinfo{journal}{Nature}
  \textbf{\bibinfo{volume}{358}}, \bibinfo{pages}{1150} (\bibinfo{year}{2009}).

\bibitem{abinitnegf}
\bibinfo{author}{\bibfnamefont{J.}~\bibnamefont{Taylor}},
  \bibinfo{author}{\bibfnamefont{H.}~\bibnamefont{Guo}}, \bibnamefont{and}
  \bibinfo{author}{\bibfnamefont{J.}~\bibnamefont{Wang}},
  \bibinfo{journal}{Phys.\ Rev.\ B} \textbf{\bibinfo{volume}{63}},
  \bibinfo{pages}{245407} (\bibinfo{year}{2001});
\bibinfo{author}{\bibfnamefont{J.~J.} \bibnamefont{Palacios}} {\it et~al.},
  \bibinfo{journal}{Phys.\ Rev.\ B} \textbf{\bibinfo{volume}{64}},
  \bibinfo{pages}{115411} (\bibinfo{year}{2001}).

\bibitem[{\citenamefont{Jacob et~al.}(2005)\citenamefont{Jacob,
  Fern\'andez-Rossier, and Palacios}}]{Jacob:prb:05}
\bibinfo{author}{\bibfnamefont{D.}~\bibnamefont{Jacob}},
  \bibinfo{author}{\bibfnamefont{J.}~\bibnamefont{Fern\'andez-Rossier}},
  \bibnamefont{and} \bibinfo{author}{\bibfnamefont{J.~J.}
  \bibnamefont{Palacios}}, \bibinfo{journal}{Phys. Rev. B}
  \textbf{\bibinfo{volume}{71}}, \bibinfo{pages}{220403(R)}
  (\bibinfo{year}{2005}).

\bibitem{nanocontacts}
\bibinfo{author}{\bibfnamefont{C.}~\bibnamefont{Untiedt}} {\it et~al.},
  \bibinfo{journal}{Phys.\ Rev.\ B} \textbf{\bibinfo{volume}{66}},
  \bibinfo{pages}{085418} (\bibinfo{year}{2002});
  \bibinfo{author}{\bibfnamefont{M.}~\bibnamefont{Viret}} {\it et~al.},
  \bibinfo{journal}{Phys.\ Rev.\ B} \textbf{\bibinfo{volume}{66}},
  \bibinfo{pages}{220401(R)} (\bibinfo{year}{2002});
\bibinfo{author}{\bibfnamefont{Z.~K.} \bibnamefont{Keane}},
  \bibinfo{author}{\bibfnamefont{L.~H.} \bibnamefont{Yu}}, \bibnamefont{and}
  \bibinfo{author}{\bibfnamefont{D.}~\bibnamefont{Natelson}},
  \bibinfo{journal}{Appl. Phys. Lett.} \textbf{\bibinfo{volume}{88}},
  \bibinfo{pages}{062514} (\bibinfo{year}{2006});
\bibinfo{author}{\bibfnamefont{K.~I.} \bibnamefont{Bolotin}} {\it et~al.},
  \bibinfo{journal}{Nano Lett.} \textbf{\bibinfo{volume}{6}},
  \bibinfo{pages}{123} (\bibinfo{year}{2006}).

\bibitem[{\citenamefont{Smogunov et~al.}(2006)\citenamefont{Smogunov,
  Dal{}Corso, and Tosatti}}]{Smogunov:prb:06}
\bibinfo{author}{\bibfnamefont{A.}~\bibnamefont{Smogunov}},
  \bibinfo{author}{\bibfnamefont{A.}~\bibnamefont{Dal{}Corso}},
  \bibnamefont{and} \bibinfo{author}{\bibfnamefont{E.}~\bibnamefont{Tosatti}},
  \bibinfo{journal}{Phys. Rev. B} \textbf{\bibinfo{volume}{73}},
  \bibinfo{pages}{075418} (\bibinfo{year}{2006}).

\bibitem[{\citenamefont{Kotliar et~al.}(2006)\citenamefont{Kotliar, Savrasov,
  Haule, Oudovenko, Parcollet, and Marianetti}}]{Kotliar:rmp:06}
\bibinfo{author}{\bibfnamefont{G.}~\bibnamefont{Kotliar}} {\it et~al.},
\bibinfo{journal}{Rev. Mod. Phys.}
\textbf{\bibinfo{volume}{78}}, \bibinfo{pages}{865} (\bibinfo{year}{2006}).

\bibitem[{\citenamefont{Thygesen and Rubio}(2007)}]{Thygesen:jcp:07}
\bibinfo{author}{\bibfnamefont{K.~S.} \bibnamefont{Thygesen}} \bibnamefont{and}
  \bibinfo{author}{\bibfnamefont{A.}~\bibnamefont{Rubio}}, \bibinfo{journal}{J
  . Chem. Phys.} \textbf{\bibinfo{volume}{126}}, \bibinfo{pages}{091101}
  (\bibinfo{year}{2007}).

\bibitem[{\citenamefont{Ferretti et~al.}(2005)\citenamefont{Ferretti,
  Calzolari, Felice, Manghi, Caldas, Nardelli, and Molinari}}]{Ferretti:prl:05}
\bibinfo{author}{\bibfnamefont{A.}~\bibnamefont{Ferretti}} {\it et~al.},
\bibinfo{journal}{Phys. Rev. Lett.} \textbf{\bibinfo{volume}{94}},
\bibinfo{pages}{116802} (\bibinfo{year}{2005}).

\bibitem[{foo()}]{epaps}
\bibinfo{note}{See EPAPS No. XXXX for details of the supercell approach, 
a detailed derivation of eq. (3), and a display of the hybridization function.}

\bibitem[{\citenamefont{Dovesi et~al.}()\citenamefont{Dovesi, Saunders, Roetti,
  Orlando, Zicovich-Wilson, Pascale, Civalleri, Doll, Harrison, Bush
  et~al.}}]{Crystal:06}
\bibinfo{author}{\bibfnamefont{R.}~\bibnamefont{Dovesi}} {\it et~al.},
\bibinfo{howpublished}{CRYSTAL06, Release 1.0.2,
  Theoretical Chemistry Group - Universita' Di Torino - Torino (Italy)}.

\bibitem[{\citenamefont{Hurley et~al.}(1986)\citenamefont{Hurley, Pacios,
  Christiansen, Ross, and Ermler}}]{Hurley:jcp:86}
\bibinfo{author}{\bibfnamefont{M.~M.} \bibnamefont{Hurley}} {\it et~al.},
  \bibinfo{journal}{J. Chem. Phys} \textbf{\bibinfo{volume}{84}},
  \bibinfo{pages}{6840} (\bibinfo{year}{1986}).

\bibitem[{\citenamefont{Grechnev et~al.}(2007)\citenamefont{Grechnev,
  Di{}Marco, Katsnelson, Lichtenstein, Wills, and Eriksson}}]{Grechnev:prb:07}
\bibinfo{author}{\bibfnamefont{A.}~\bibnamefont{Grechnev}} {\it et~al.},
  \bibinfo{journal}{Phys. Rev. B} \textbf{\bibinfo{volume}{76}},
  \bibinfo{pages}{035107} (\bibinfo{year}{2007}).

\bibitem[{\citenamefont{Haule et~al.}(2001)\citenamefont{Haule, Kirchner,
  Kroha, and W\"olfle}}]{Haule:prb:01}
\bibinfo{author}{\bibfnamefont{K.}~\bibnamefont{Haule}} {\it et~al.},
  \bibinfo{journal}{Phys. Rev. B} \textbf{\bibinfo{volume}{64}},
  \bibinfo{pages}{155111} (\bibinfo{year}{2001}).

\bibitem[{\citenamefont{Meir and Wingreen}(1992)}]{Meir:prl:92}
\bibinfo{author}{\bibfnamefont{Y.}~\bibnamefont{Meir}} \bibnamefont{and}
  \bibinfo{author}{\bibfnamefont{N.~S.}~\bibnamefont{Wingreen}},
  \bibinfo{journal}{Phys.\ Rev.\ Lett.} \textbf{\bibinfo{volume}{68}},
  \bibinfo{pages}{2512} (\bibinfo{year}{1992}).

\bibitem[{\citenamefont{Landauer}(1970)}]{Landauer:philmag:70}
\bibinfo{author}{\bibfnamefont{R.}~\bibnamefont{Landauer}},
  \bibinfo{journal}{Philos. Mag.} \textbf{\bibinfo{volume}{21}},
  \bibinfo{pages}{863} (\bibinfo{year}{1970}).

\bibitem[{foo()}]{foot1}
\bibinfo{note}{For the idealized contact geometry considered here, the voltage drop would 
actually be symmetric. However, in reality the geometries are often quite asymmetric so that 
our assumption is justified.}

\bibitem[{\citenamefont{Hewson}(1993)}]{Hewson:book}
\bibinfo{author}{\bibfnamefont{A.~C.} \bibnamefont{Hewson}},
  \emph{\bibinfo{title}{The Kondo problem to heavy fermions}}
  (\bibinfo{publisher}{Cambridge University Press}, \bibinfo{year}{1993}).

\bibitem[{\citenamefont{Jamneala et~al.}(2000)\citenamefont{Jamneala, Madhavan,
  Chen, and Crommie}}]{Jamneala:prb:00}
\bibinfo{author}{\bibfnamefont{T.}~\bibnamefont{Jamneala}} {\it et~al.},
  \bibinfo{journal}{Phys. Rev. B} \textbf{\bibinfo{volume}{61}},
  \bibinfo{pages}{9990} (\bibinfo{year}{2000}).

\end{thebibliography}

\end{document}